\documentclass[10pt,prd,twocolumn,preprintnumbers,amsmath,amssymb]{revtex4-1}

\usepackage{import}
\usepackage{hyphenat}
\usepackage{array}
\usepackage[british]{babel}
\usepackage{graphicx}
\usepackage{color}			
\usepackage{epstopdf}
\usepackage{pict2e}			
\usepackage{empheq}
\usepackage{bbm}

\providecommand{\D}{\textnormal{d}}
\providecommand{\Hz}{\textnormal{Hz}}

\usepackage[colorlinks]{hyperref}

\begin{document}
\title{Imprints of the QCD Phase Transition on the Spectrum of Gravitational Waves}
\author{Simon Schettler}
\affiliation{Institut f\"ur Theoretische Physik, Universit\"at Heidelberg\\ 
Philosphenweg 16, D-69120 Heidelberg, Germany}
\author{Tillmann Boeckel}
\affiliation{Institut f\"ur Theoretische Physik, Universit\"at Heidelberg\\ 
Philosphenweg 16, D-69120 Heidelberg, Germany}
\author{J\"urgen Schaffner-Bielich}
\affiliation{Institut f\"ur Theoretische Physik, Universit\"at Heidelberg\\ 
Philosphenweg 16, D-69120 Heidelberg, Germany}
\date{\today}

\begin{abstract}    
We have investigated effects of the QCD phase transition on the relic GW spectrum applying several equations of state for the strongly interacting matter: Besides the bag model, which describes a first order transition, we use recent data from lattice calculations featuring a crossover. Finally, we include a short period of inflation during the transition which allows for a first order phase transition at finite baryon density. 
Our results show that the QCD transition imprints a step into the spectrum of GWs. 
Within the first two scenarios, entropy conservation leads to a step-size determined by the relativistic degrees of freedom before and after the transition. 
The inflation of the third scenario much stronger attenuates the high-frequency modes: 
An inflationary model being consistent with observation entails suppression of the spectral energy density by a factor of $\lesssim 10^{-12}$.
\end{abstract}

\maketitle

\selectlanguage{british}
Theories of inflation predict the existence of a background spectrum of gravitational waves (GWs) which has been created in the very early universe together with scalar perturbations of the energy density. 
Since then the GWs propagate freely through spacetime still retaining information about the conditions under which they originated. This is because their cross section is very small: Particles with larger cross section, such as neutrinos or photons, decouple much later and therefore carry no information about the universe at such early times and high energies. This advantage of GWs renders their direct detection impossible until today. 
However, observations of the binary pulsar B1913+16 (Hulse-Taylor pulsar) provide strong evidence for energy loss through emission of gravitational radiation \cite{2005ASPC..328...25W}.

In this article we demonstrate that in spite of being decoupled, relic GWs can show imprints of subsequent cosmic events, e.g. phase transitions. 
We present the shape of the relic GW spectrum after different scenarios of the QCD phase transition. Before the transition, at about $200\,$MeV, the universe is filled with a fluid containing relativistic quarks (up, down and strange), gluons, leptons and photons. After the transition, quarks are confined into hadrons of which the pions are the only degrees of freedom which are still relativistic. 
From this reduction of degrees of freedom and the assumption of entropy conservation, 
 the impact of the phase transition on the spectrum of GWs can be derived. Numerical calculations show that the shape of the spectrum after confinement and chiral symmetry breaking does not strongly depend on the order of the phase transition.

However, there are scenarios for the cosmological QCD phase transition where entropy conservation is considerably violated and the gross features of the spectrum are not fixed by the estimates mentioned above. We will discuss a model which includes a short period with dominating vacuum energy density causing inflation. The melting of this energy is associated with a large entropy release. Within this model, the energy density in high-frequency GWs is much more diluted than within the scenarios where entropy is approximately conserved.

We emphasize that the relic GWs, generated long time before the QCD transition, are not the only ones which are of interest in our context: During a first order transition, bubbles of the new phase form within the old phase and grow under certain conditions until the fluid is completely converted into the new phase. This process involves bubble collisions and turbulences which are a source of gravitational radiation. 
For a standard first order transition the resulting GW spectra are already explored (see, for example, \cite{2008JCAP...09..022H,PhysRevD.49.2837,PhysRevD.82.063511} and references therein), 
but a careful investigation within an inflationary QCD transition is still to be done.

The structure of this article is as follows: After having collected a few important equations in the next two sections, we describe a model which allows for a first order QCD phase transition being consistent with both observation and lattice data (Section \ref{s:Model}). In Section \ref{s:Calculations} some analytic estimates and the results of the numerical calculations are presented. The calculations have been done within the bag model, using various parameterizations of lattice data, and within the inflationary scenario. We conclude in Section \ref{s:Conclusions}.

\section{Gravitational Waves within a FRW Background}

In the very early universe, a period of inflation is assumed to have taken place. This leads to perturbations within a homogeneous background. The tensorial part of these perturbations consists of GWs which are described by the metric
\begin{equation*}
\label{GWmetric}
	ds^2 = a^2 \left(\D \eta^2 - \left(\delta_{ij} - h_{ij}\right) \D x^i\D x^j  \right).
\end{equation*}
Einstein's equation for this metric and an energy-momentum tensor of an ideal fluid gives the equation of motion for a GW within a Friedmann-Robertson-Walker background:
\begin{equation}
\label{EOMofGWfull}
	h''_{ij} + 2\mathcal{H}h'_{ij} - \Delta h_{ij} = 0.
\end{equation}
$\mathcal{H} = a'/a$ is the conformal Hubble parameter and primes denote derivatives with respect to conformal time.
Now the expansion
\begin{equation}
\label{hExpansion}
	h_{ij} (\eta, {\bf x}) = \int \frac{d^3k}{(2\pi)^{3/2}} e_{ij} ({\bf k})h_{\bf k}(\eta) e^{i {\bf k} \cdot {\bf x}},
\end{equation}
with $e_{ij}({\bf k})$ being the polarization tensor of the mode {\bf k}, 
is applied to \eqref{EOMofGWfull}:
\begin{equation*}
	h_{\bf k}''(\eta) + 2\mathcal{H} h'_{\bf k}(\eta) + k^2 h_{\bf k}(\eta) = 0.
\end{equation*}
With $v_{\bf k}(\eta):= a h_{\bf k}(\eta)$ we get
\begin{equation}
\label{EOMofGW}	
	v''_{\bf k} + \left(k^2 - \frac{a''}{a}\right)v_{\bf k} = 0
\end{equation}
(see \cite{Maggiore:1999vm} for comparison).
We solve this equation in order to calculate the behavior of the GW spectrum during the QCD phase transition.  Therefore we need $a''/a$ as a function of time or temperature:
\begin{equation*}
	\frac{a''}{a} = a^2 \left(H^2 + \frac{\ddot{a}}{a}\right) = \frac{ 4\pi G}{3} a^2 \left(\rho - 3 p \right),
\end{equation*}
where the Friedmann equations have been used. We see that for our calculations we just have to know the evolution of both the scale factor $a$ and the trace of the energy momentum tensor.

\section{The Bag Model}

A description of confinement and asymptotic freedom is given by the bag model \cite{PhysRevD.9.3471} which assigns a positive, constant energy density $B$ to the interior of hadrons. 
At very high densities or temperatures, the hadrons might overlap forming a single bag in which the quarks and gluons can move almost freely. In this way, the bag model can describe the confinement-deconfinement phase transition of strongly interacting matter.

Assuming a gas of freely moving relativistic quarks and gluons above a critical temperature $T_ \textnormal{c} $ and a gas of freely moving relativistic hadrons below $T_ \textnormal{c} $, one can write down the entropy density for both phases:
\begin{equation*}
	s(T) = \frac{2 \pi^2}{45} g(T) T^3 \quad \text{with} \quad g(T) := \begin{cases} 
		g_1 \quad\text{if} \;\, T \geq T_ \textnormal{c} \\
		g_2 \quad\text{if} \;\, T < T_ \textnormal{c}.
		\end{cases}
\end{equation*}
$g_1$ and $g_2$ are the numbers of effective relativistic degrees of freedom. For $g_1= 51.25$ we take into account gluons, up and down quarks, photons, electrons, positrons, \mbox{(anti-)}muons and neutrinos; $g_2=17.25$ only includes pions instead of quarks and gluons. The Maxwell relation
\begin{equation*}
	\left( \frac{\partial S}{\partial V} \right)_{T,N} =\left( \frac{\partial p}{\partial T} \right)_{V,N} 
\end{equation*}
yields
\pagebreak
\begin{align}
\label{BagPressure}
	p(T) = \int s(T) dT &= \begin{cases} 
 			  \frac{\pi^2}{90} g_1 T^4 - B &\text{if} \;T \geq T_ \textnormal{c} \\
  			\frac{\pi^2}{90} g_2 T^4 &\text{if} \;T < T_ \textnormal{c},
  		\end{cases}\\
\label{BagEnergy}	
	\rho(T) = sT - p &=
 		\begin{cases} 
 			\frac{\pi^2}{30} g_1 T^4 + B &\text{if} \;T \geq T_ \textnormal{c} \\
  			\frac{\pi^2}{30} g_2 T^4 &\text{if} \;T < T_ \textnormal{c},
  		\end{cases}
\end{align}
where the bag constant is fixed by $T_ \textnormal{c} $ and the difference in the degrees of freedom in both phases:
\begin{equation*}
	B = \frac{\pi^2}{90} (g_1 - g_2)T_ \textnormal{c} ^4.
\end{equation*}

\section{An Inflationary QCD Phase Transition}\label{s:Model}

\begin{figure}
\setlength{\unitlength}{1cm}
\resizebox{8.5cm}{!}
	{
		\includegraphics{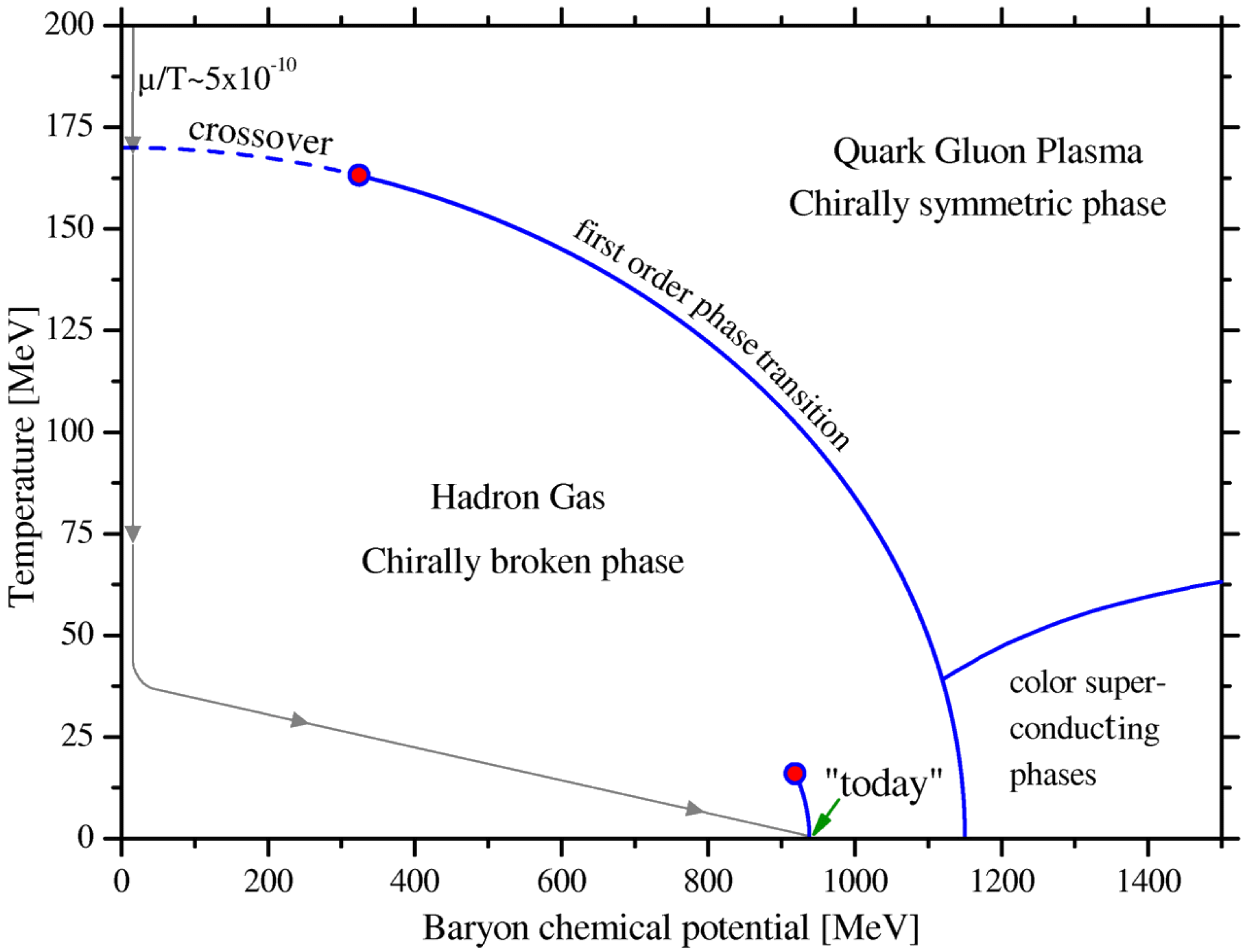}
	}	
\caption{(Color online) Sketch of a possible phase diagram for strongly interacting matter. The early universe is generally assumed to evolve at very small baryon chemical potential, where lattice calculations predict a cross-over. At higher $\mu_ \textnormal{b} $ a first order phase transition is possible. The evolution of the universe is indicated by arrows, see also \cite{2002astro.ph.11346F}. 
\label{figQCDPhaseDiagram} 
}
\end{figure}

Lattice QCD calculations show a crossover between the hadronic phase and the quark-gluon plasma at temperatures of about 150-200 MeV. However, until now LQCD makes reliable predictions only for small chemical potential. For example, at chemical potentials of order $\mu_ \textnormal{b} \sim T$ the character of the transition between the two phases is unclear. In principle, a first order phase transition is possible in this region of the QCD phase diagram. 

Within the standard model of cosmology the baryon chemical potential during QCD time is expected to be well within the cross-over region of the phase structure, leading to the assumption that a cosmological first order transition is unlikely (see Figure~\ref{figQCDPhaseDiagram}). More precisely, the baryon asymmetry calculated from CMB and BBN data is $\eta_ \textnormal{b} \sim 6\cdot 10^{-10}$ \cite{2003PhLB..566....8B}, and for small values of $\eta_\textnormal{b}$ the thermodynamics of a relativistic Fermi gas gives
\begin{equation*}
	\eta_\textnormal{b} \approx \frac{5g_\textnormal{q}}{7\pi^2g}\frac{\mu_\textnormal{b}}{T},
\end{equation*}
where $g$ is the effective number of helicity states and in $g_\textnormal{q}$ only quarks are accounted for.
There are nevertheless different ways to circumvent this scenario and to preserve a first order transition: It has been pointed out in \cite{2009JCAP...11..025S} that a high lepton asymmetry could be sufficient. Another possibility is considered in \cite{2010PhRvL.105d1301B}: If a short period of inflation occurs during the QCD phase transition, the baryon density could be diluted such that the corresponding chemical potential starts at values $\mu(T_\textnormal{QCD} ) \sim T_ \textnormal{QCD} $ (allowing the transition to be first order) and nonetheless matches the observations today. For example, the high baryon density could be provided by an Affleck-Dine baryogenesis before QCD time \cite{2003RvMP...76....1D}. The possibility of an inflationary QCD phase transition has also been considered in \cite{1990AN....311..265B,1990ZPhyC..48..147J}. \cite{2000JPhG...26..771B} discusses a transition with at least strong supercooling.

\begin{figure}[t]
\setlength{\unitlength}{1cm}

\resizebox{8.5cm}{!}
	{
		\includegraphics{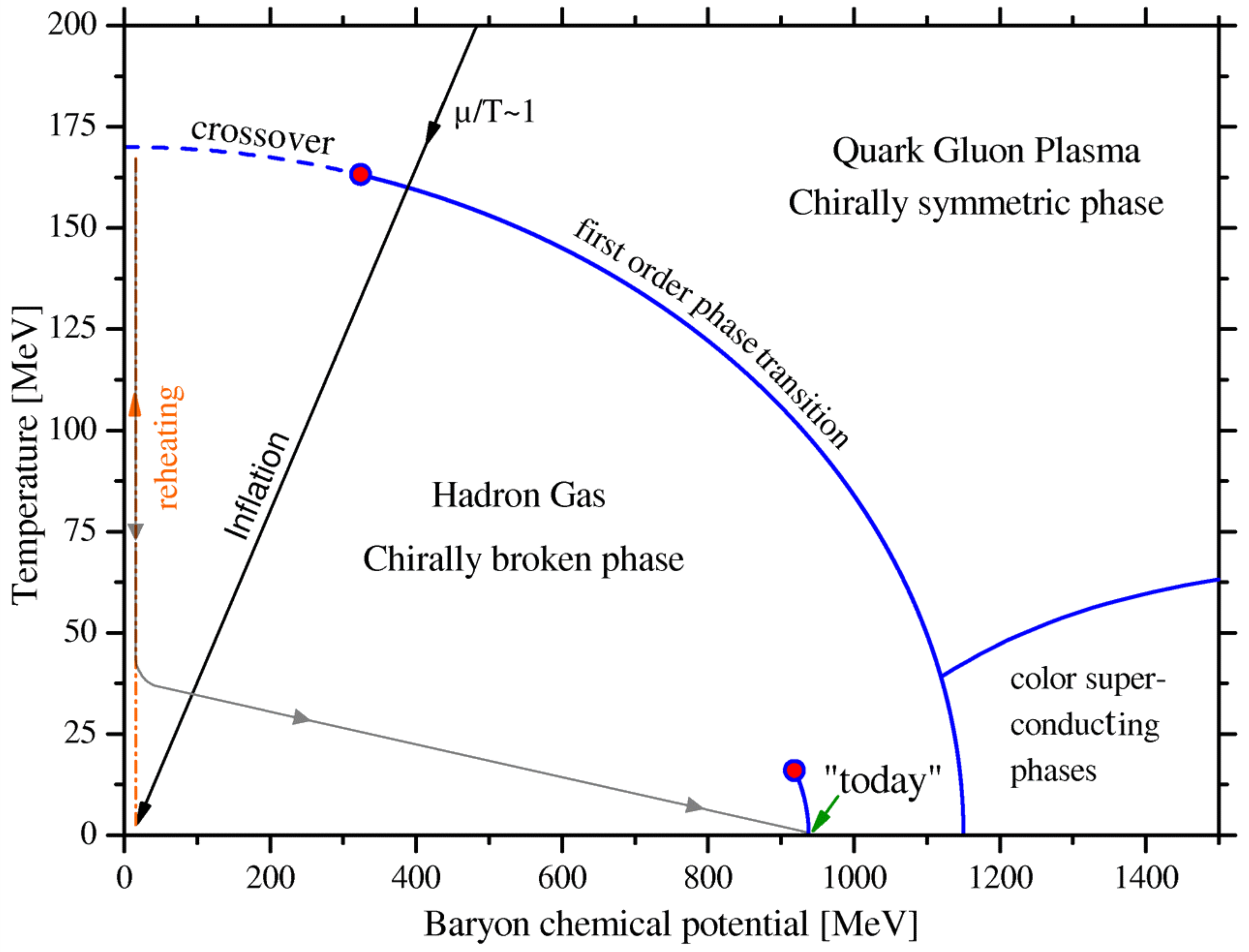}
	}
\caption{(Color online) Possible track of the background fluid through the phase diagram for strongly interacting matter. Within this scenario the system intersects the first order line starting from high chemical potential. It is supercooled during inflation after which the actual phase transition takes place and reheating sets in.  Afterwards, the evolution follows the commonly accepted path.}
\label{figInflQCDPhaseDiagram}
\end{figure}

We will be concerned with the effect of such an inflation on the relic GW spectrum. We only mention that also structure formation is influenced, primordial magnetic fields can be produced, and new GWs can be generated through bubble collisions and turbulences,  \cite{2010JPhG...37i4005B, 2010PhRvL.105d1301B}. 
Thus, the scenario describes a fluid with high baryon asymmetry (see Figure \ref{figInflQCDPhaseDiagram}) which is cooled down and reaches a quasi stable state still having a considerable amount of potential energy compared to the ground state. This potential energy plays the same role as the bag constant $B$ in the bag model. However, if the transition into the true vacuum state occurs late, the universe will be dominated by a vacuum energy leading to inflation. Baryon density and thermal energy density are both strongly diluted. Then the vacuum energy melts and energy conservation gives the temperature after reheating.  Afterwards, the fluid returns to the conventional evolution.
We will implement this scenario in a simple way, not requiring a field theoretical description.

\section{Calculations and Results}\label{s:Calculations}

\subsection{Horizon Entry}
\begin{figure}[t]
\resizebox{8.6cm}{!}
{
	\includegraphics{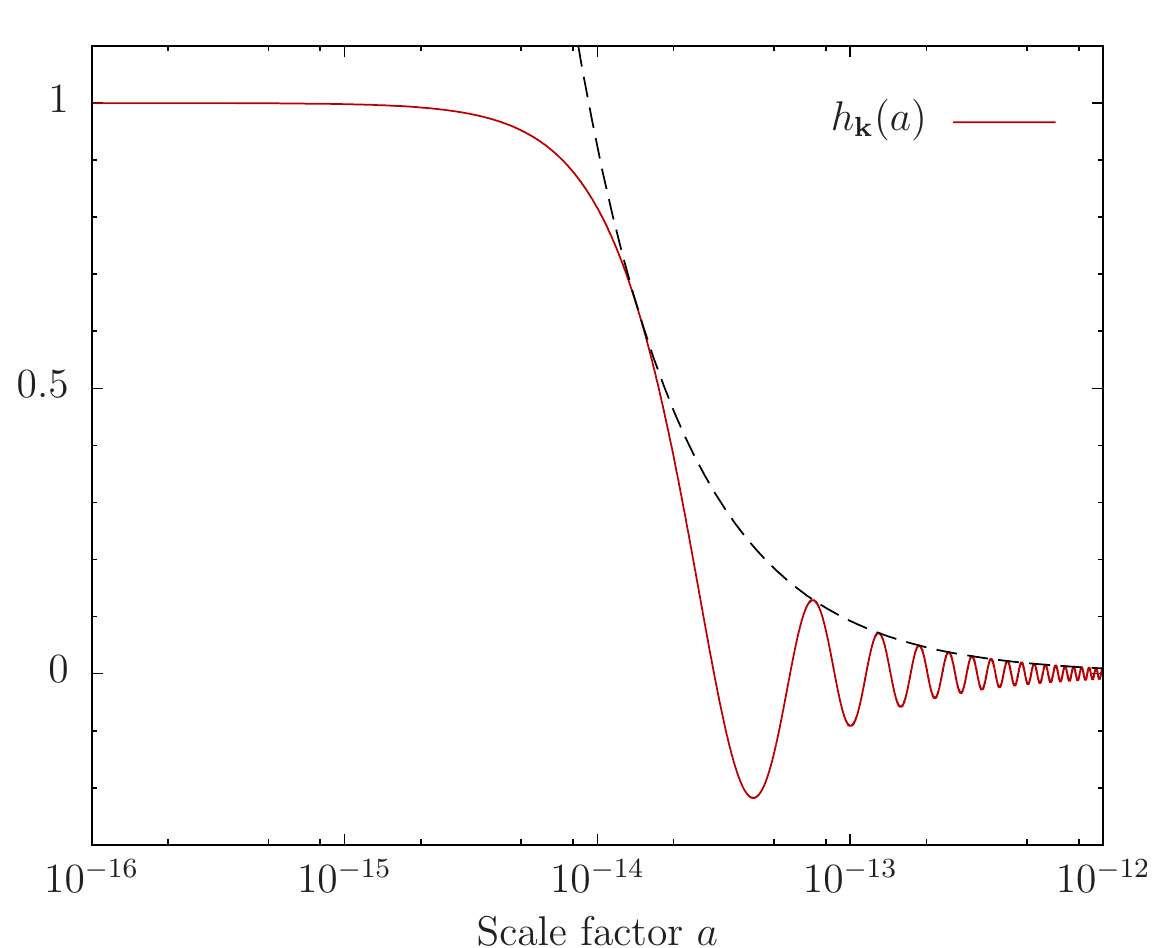}
}
\caption{(Color online) Behavior of a GW entering the horizon. For this calculation the background was assumed to be purely radiative. A phase transition is not included. The frequency of the mode was chosen to be the critical frequency: $|{\bf k}| = 2\pi \nu^*.$ The dashed line indicates a function proportional to $1/a$ fit to the amplitude of the GW after horizon entry.}
\label{figModePureRad}	
\end{figure}

Superhorizon modes ($k\eta \ll 1$) of relic gravitational radiation have a flat spectrum:
\begin{equation}
\label{PsuperKindep}
	\mathcal{P}_ \textnormal{g} ^ \textnormal{super} (k) \propto |h_{\bf k}|^2 k^3 \propto k^0.
\end{equation}
This holds true independently of changes in the equation of state. However, if the universe is not inflating, modes which have exited the horizon during inflation reenter the horizon again. This leads to a $k$-dependent power spectrum because horizon entry occurs at different times for different modes and inside the horizon the modes start decaying: $h \propto 1/a$. 
For a radiation dominated universe, we have
\begin{equation*}
	\label{GWinRaddom}
	a \propto \eta \quad \xRightarrow{\eqref{EOMofGW}} v \propto \exp(\pm \mathrm{i} k \eta).
\end{equation*}
(see Figure \ref{figModePureRad}). We denote the scale factor at horizon entry of the mode $k$ with $a_ \textnormal{in}(k)$ and see:
\begin{equation*}
	a_ \textnormal{in}(k) = \frac{k}{H_ \textnormal{in}(k)} \propto k \, a^2_ \textnormal{in}(k) \quad \Longrightarrow \quad a_ \textnormal{in}(k) \propto \frac{1}{k}.
\end{equation*}
Therefore, the power spectrum after horizon entry is
\begin{equation}
\label{PSpectHorEntry}
	\mathcal{P}_ \textnormal{g} ^ \textnormal{sub} (k) = \mathcal{P}_ \textnormal{g} ^ \textnormal{super} (k) \left( \frac{a_ \textnormal{in}(k)}{a}\right)^2 \propto \frac{1}{k^2}.
\end{equation}
\noindent Inserting the expansion \eqref{hExpansion} into the expression for the energy density of gravitational waves \cite{Maggiore:1999vm},
\begin{equation*}
	\rho_\textnormal{g} = \frac{1}{32\pi G} \bigl< (\partial_0 h_{ij})(\partial_0 h^{ij}) \bigr>,
\end{equation*}
we get the $k$-dependence of their energy fraction per logarithmic frequency interval:
\begin{equation}
\label{OmegaPRelation}
	\Omega_\textnormal{g}(k) := \frac{1}{\rho_ \textnormal{crit}} \frac{d\rho_\textnormal{g}}{d\ln k} \propto k^5 |h_{\bf k}|^2 \propto k^2 \mathcal{P}_ \textnormal{g}(k).
\end{equation}
$\Omega_\textnormal{g}$ does not depend on $k$ after horizon entry during radiation domination.

\subsection{A Step in the Spectrum}
The QCD phase transition imprints a step into the primordial differential energy spectrum of GWs $\Omega_ \textnormal{g} (k)$. Let us first estimate some characterizing properties of this step. For comparison see the discussion in \cite{1998MPLA...13.2771S}.\\

\noindent {\bf The typical frequency scale.} We calculate the frequency of modes which enter the horizon at the end of the transition.  The Hubble parameter for the critical temperature is
\begin{equation*}
H_ \textnormal{c} ^2 = \frac{8\pi G}{3} \rho (T_ \textnormal{c} ) = \frac{8\pi G}{3} \frac{\pi^2}{30} g_2 T_ \textnormal{c} ^4,
\end{equation*}
which corresponds to a comoving wave number $k^*$ with
\begin{equation*}
	\frac{a_ \textnormal{c} H_ \textnormal{c} }{k^*} \approx 1\quad \Longrightarrow \quad \frac{k^*}{a_ \textnormal{c} } =\sqrt{\frac{8 \pi^3 G}{90}g_2} \, T_ \textnormal{c} ^2.
\end{equation*}
In order to calculate the corresponding frequency today we need the ratio of the scale factors today ($a_0$) and at QCD time ($a_ \textnormal{c} $):
\begin{equation*}
	\frac{a_ \textnormal{c} }{a_0} = \frac{T_0}{T_ \textnormal{c} } \left(\frac{g_0}{g_2}\right)^{1/3}.
\end{equation*}
So for the physical wave number today, $k_0^*$, we obtain
 \begin{equation*}
	k_0^* = \frac{k^*}{a_ \textnormal{c} }  \frac{T_0}{T_ \textnormal{c} } \left(\frac{g_0}{g_2}\right)^{1/3}.
\end{equation*}
Then the frequency today is:
\begin{equation*}
\nu^* = 3.53 \frac{T_ \textnormal{c} }{180\, \textnormal{MeV} } \left(\frac{g_2}{17.25}\right)^{1/6} 10^{-9}\, \Hz.\notag
\end{equation*}
For this result the values $T_{0} = 2.35\cdot 10^{-4}\,$eV and $g_0 = 3.91$ have been inserted. In spite of being nonrelativistic today the neutrinos contribute to $g_0$ because they do not transfer their entropy to the photons after decoupling. Thus, there is no shift in the standard redshift relation $T \propto a^{-1}$.\\

\noindent {\bf The step-size.} In order to show that $\Omega_ \textnormal{g}$ is damped for high frequencies by a factor of 0.7, we collect the information of equations \eqref{PsuperKindep}, \eqref{PSpectHorEntry} and \eqref{OmegaPRelation} to obtain for the $k$-dependence of $\Omega_ \textnormal{g} $
\begin{equation*}
	\Omega_ \textnormal{g} (k) \propto k^2a_ \textnormal{in} ^2(k).
\end{equation*}
Next we use entropy conservation and get for radiation dominated periods:
\begin{equation*}
	H \propto T^2 \sqrt{g} \propto \frac{\sqrt g}{g^{2/3} a^2} = g^{-1/6} a^{-2},
\end{equation*}
and replace $k$ by the Hubble length at horizon entry ($k = H_ \textnormal{in} (k) a_ \textnormal{in} (k)$):
\begin{equation*}
	\Omega_ \textnormal{g} (k) \propto H_ \textnormal{in} ^2 a_ \textnormal{in} ^4 \propto g_k^{-1/3}.
\end{equation*}
With $g_k$ we denote the number of relativistic degrees of freedom at horizon entry of the mode $k$. Assuming a radiation dominated universe well before and after the QCD phase transition leads to
\begin{equation*}
	\frac{\Omega_ \textnormal{g}  (\nu \gg \nu^*)}{\Omega_ \textnormal{g}  (\nu \ll \nu^*)} = \left( \frac{g_2}{g_1} \right) ^{1/3} \approx 0.696.
\end{equation*}
\vspace{0.5cm}

\noindent{\bf The slope of the step.} Let us revisit the question of how the energy spectrum looks like after horizon entry. We will do the calculation for an arbitrary equation of state $p = w \rho$ with the only restriction that $w(a)$ varies slowly with $a$. We start from energy-momentum conservation,
\begin{gather}
	\label{dlnHdlna}
	 \D \ln H(a) = -\frac{3}{2}(1 + w(a)) \D \ln a.
\end{gather}
From this we get
\begin{align*}	
	\D \ln \nu = \D \ln (H_ \textnormal{in}  a_ \textnormal{in} ) = \left[ -\frac{1}{2}(1 + 3w(a))\D \ln a \right]_ \textnormal{in} 
\end{align*}
 which we use in
 \begin{align*}
 	\frac{d\ln\Omega_ \textnormal{g} (\nu)}{d\ln \nu} =\left[\frac{1}{-\frac{1}{2}(1+3w)} \frac{d\ln k^2 a^2}{d\ln a}\right]_ \textnormal{in} .
 \end{align*}
 We write
 \begin{equation}
 \label{ainAndk}
 	[k^2a^2]_ \textnormal{in}  = [H^2a^4]_ \textnormal{in}  \propto [a^{-3(1+w)}a^4]_ \textnormal{in} ,
 \end{equation}
 noting that the proportionality is only strict if $w$ in \eqref{dlnHdlna} does not depend on $a$. So
 \begin{equation}
 \label{OmegaAndW}
 	\frac{d\ln\Omega(\nu)}{d\ln \nu} = -2 \frac{1-3w(a_ \textnormal{in} )}{1+3w(a_ \textnormal{in} )}.
 \end{equation}

\subsection{Bag Model}

\begin{figure}[t]
	\resizebox{8.8cm}{!}{	\includegraphics{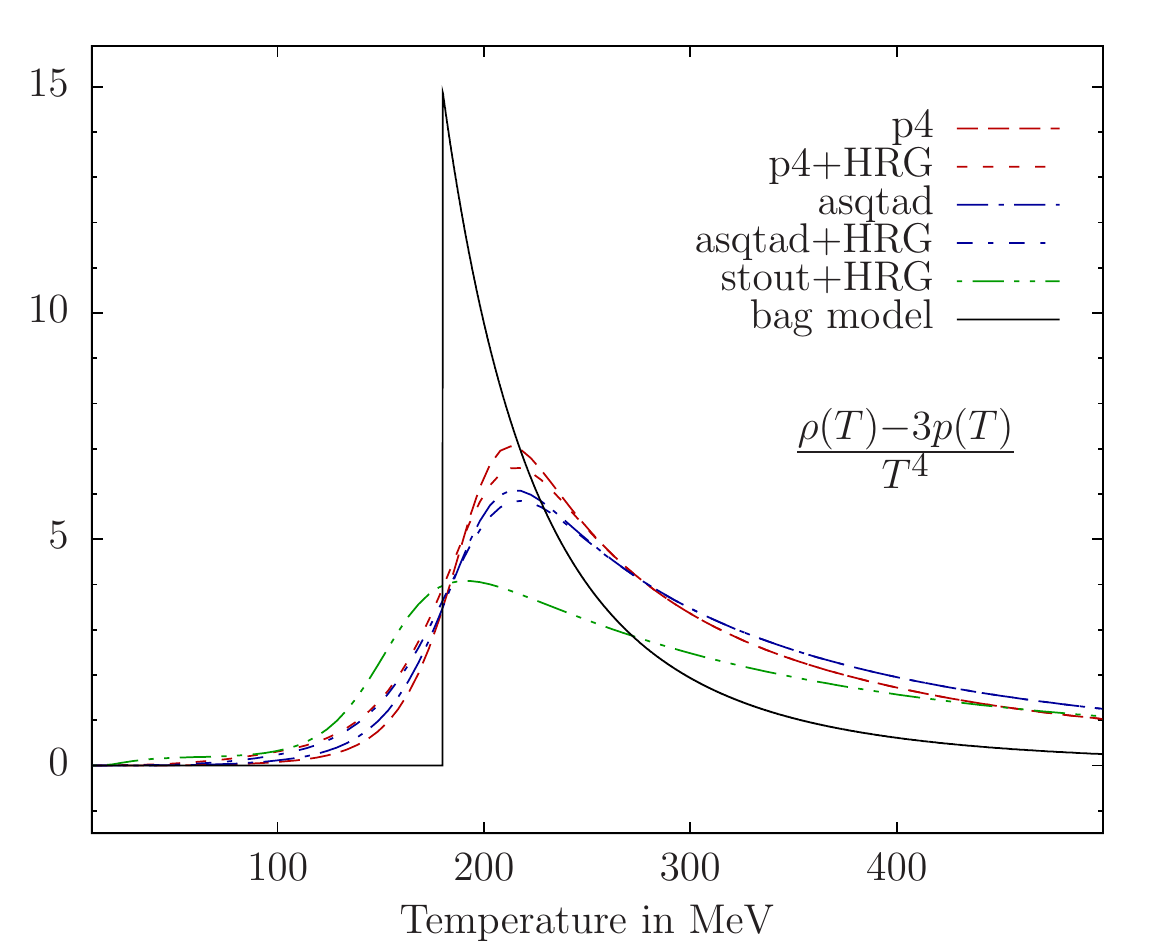}}
	\vspace{-0.5cm}
	\caption{(Color online) The trace anomaly of strongly interacting matter, normalized to the temperature. Within the bag model, the only contribution comes from the bag constant $B$. The dashed curves depict parameterizations of results of lattice calculations. More explanation is provided in the text.}
	\label{fig:TraceLattice}
\end{figure}

To keep things as simple as possible we first employ the bag model for a numerical calculation. So we assume a pressure and energy density according to equations \eqref{BagPressure} and \eqref{BagEnergy}, which leads to 
\begin{equation*}
	\rho(T) - 3p(T) = 
	\begin{cases}
		4B 	&\text{if}\;T \geq T_ \textnormal{c} \\
	 	0 	&\text{if}\; T < T_ \textnormal{c} 
	\end{cases}
\end{equation*}
(see Figure \ref{fig:TraceLattice}). Note that within this setup, we do not account for supercooling: We assume a phase transition beginning exactly at $T = T_ \textnormal{c} $. Correspondingly, we assume that entropy is conserved during the transition. 

The corresponding calculation has already been done in \cite{1998MPLA...13.2771S}.
Our result can be seen in Figure~\ref{fig:SpectrumLattice}. As expected from our estimates, the energy density of relic gravitational waves is reduced by a factor $\sim0.7$ for modes with frequencies higher than the critical frequency scale $\nu^*$.

\subsection{Lattice Data}
In order to build on a more realistic description of strongly interacting matter we take into account results from LQCD in this section. In particular, we use data published by the Bielefeld-BNL/RIKEN-Columbia collaboration in \cite{PhysRevD.80.014504} which are based on calculations including three quarks ($u$, $d$, and $s$) with physical strange quark mass and two degenerate light quark masses being one tenth of the strange quark mass. The equation of state has been calculated for two different improved staggered fermion actions, the asqtad and p4 actions. They are both $\mathcal O(a^2)$ where $a$ is the lattice spacing. 
The published results are mostly extracted from calculations on lattices of size $32^3 \times 8$ where $N=32$ is the spatial extent and $N_\tau = 8$ is the temporal extent. We have also calculated with data published by the Wuppertal-Budapest collaboration in \cite{2010arXiv1005.3508B, 2010arXiv1007.2580B} which are obtained using a different type of staggered fermion action, the stout action.

The only things we use as an input in our calculations are the parameterizations of the trace anomaly given in \cite{PhysRevD.80.014504,2010arXiv1007.2580B}. In order to better account for the low temperature physics, data from {\it hadron resonance gas} calculations have been incorporated over a range of temperatures 100~MeV $< T <$ 130 MeV.  They slightly reduce the peak in the transition region. The label ``HRG'' is given to the quantities resulting from this procedure. In the ``stout'' case, the parameterization is chosen such that it also reflects the HRG results  in this region.
\begin{figure}[t]
	\resizebox{8.8cm}{!}{	\includegraphics{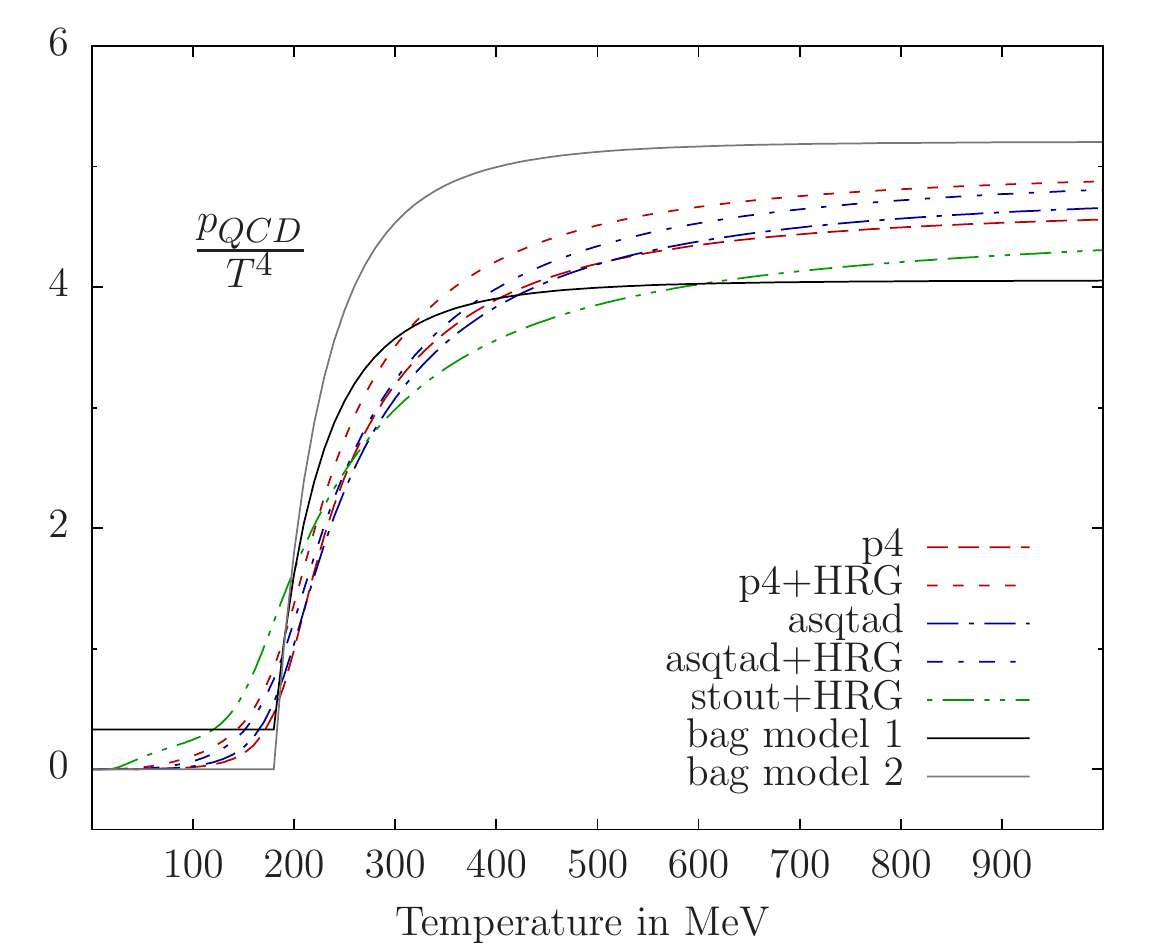}}
	\vspace{-0.5cm}
	\caption{(Color online) The pressure of strongly interacting matter, normalized to the temperature. The dashed lines correspond to lattice data. For bag model 1 we included pions in the low temperature phase and we excluded strange quarks from the high temperature phase: $g_ \textnormal{quarks} = 37$, $g_ \textnormal{hadrons} = 3$. For bag model 2 it is the opposite way around: $g_ \textnormal{quarks} = 47.5$, $g_ \textnormal{hadrons} = 0$. The lattice results tend to the Stefan-Boltzmann limit of bag model 2.}
	\label{fig:PressureLattice}
\end{figure}

\begin{figure}[t]
	\resizebox{8.8cm}{!}{	\includegraphics{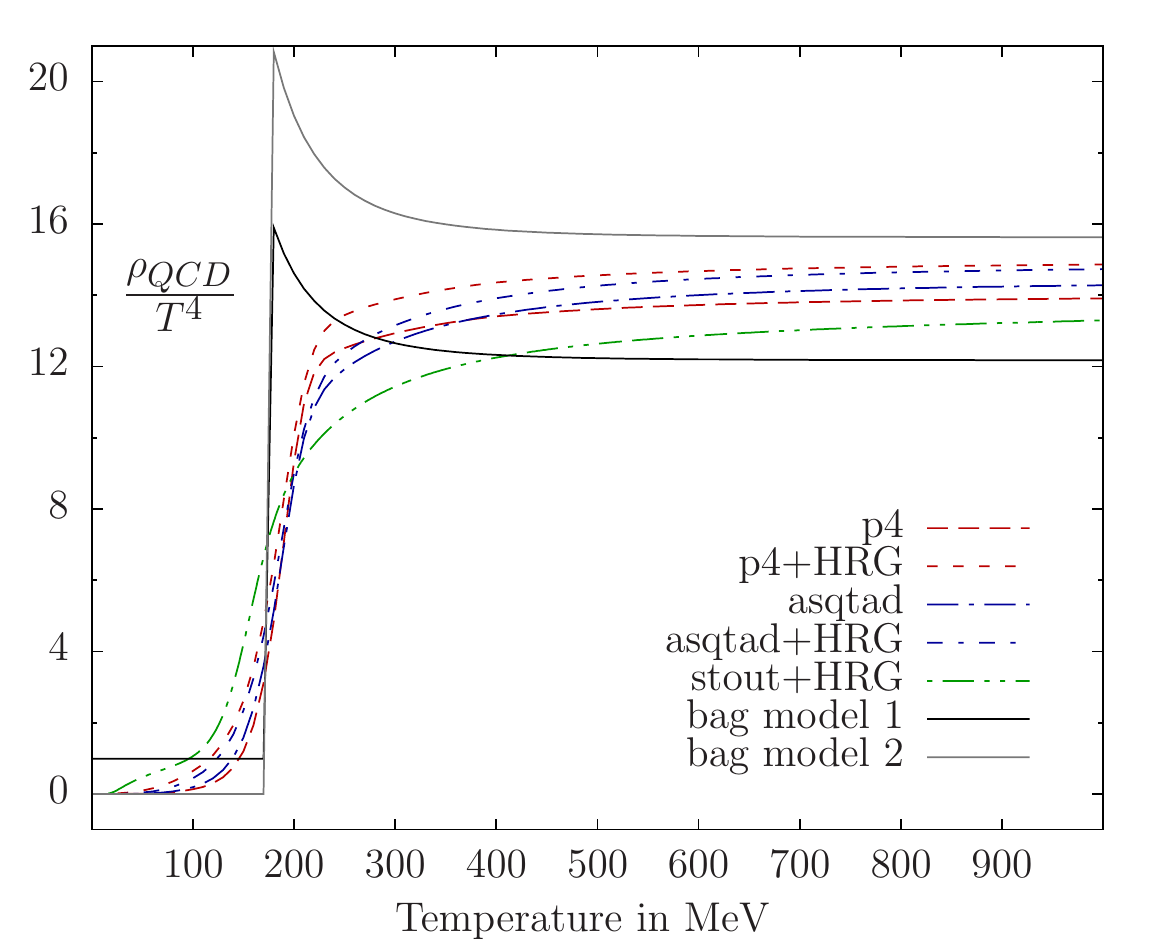}}
	\vspace{-0.5cm}
	\caption{(Color online) The energy density of strongly interacting matter within the bag model and according to lattice data. More information is provided in the caption of Figure \ref{fig:PressureLattice}.}
	\label{fig:RhoLattice}
\end{figure}

The pressure of the medium is given by
\begin{equation*}
	\frac{p(T)}{T^4} - \frac{p(T_0)}{T_0^4} = \int^T_{T_0} dT' \frac{\rho - 3p}{T'^5}
\end{equation*}
which, together with the parameterizations of the trace anomaly, also fixes the energy density.
The temperature dependence of both quantities is shown in Figures \ref{fig:PressureLattice} and \ref{fig:RhoLattice}.

Knowing the energy density and the trace anomaly as functions of temperature we can now calculate the spectrum of relic GWs after a QCD transition described by lattice data. The result displayed in Figure \ref{fig:SpectrumLattice} looks very similar to the one we have found within the bag model: In both calculations we find a step at similar frequencies with almost the same step-size, which is what we have expected. The step-size is not exactly the same: 
\begin{figure}[t]
	\resizebox{8.8cm}{!}{	\includegraphics{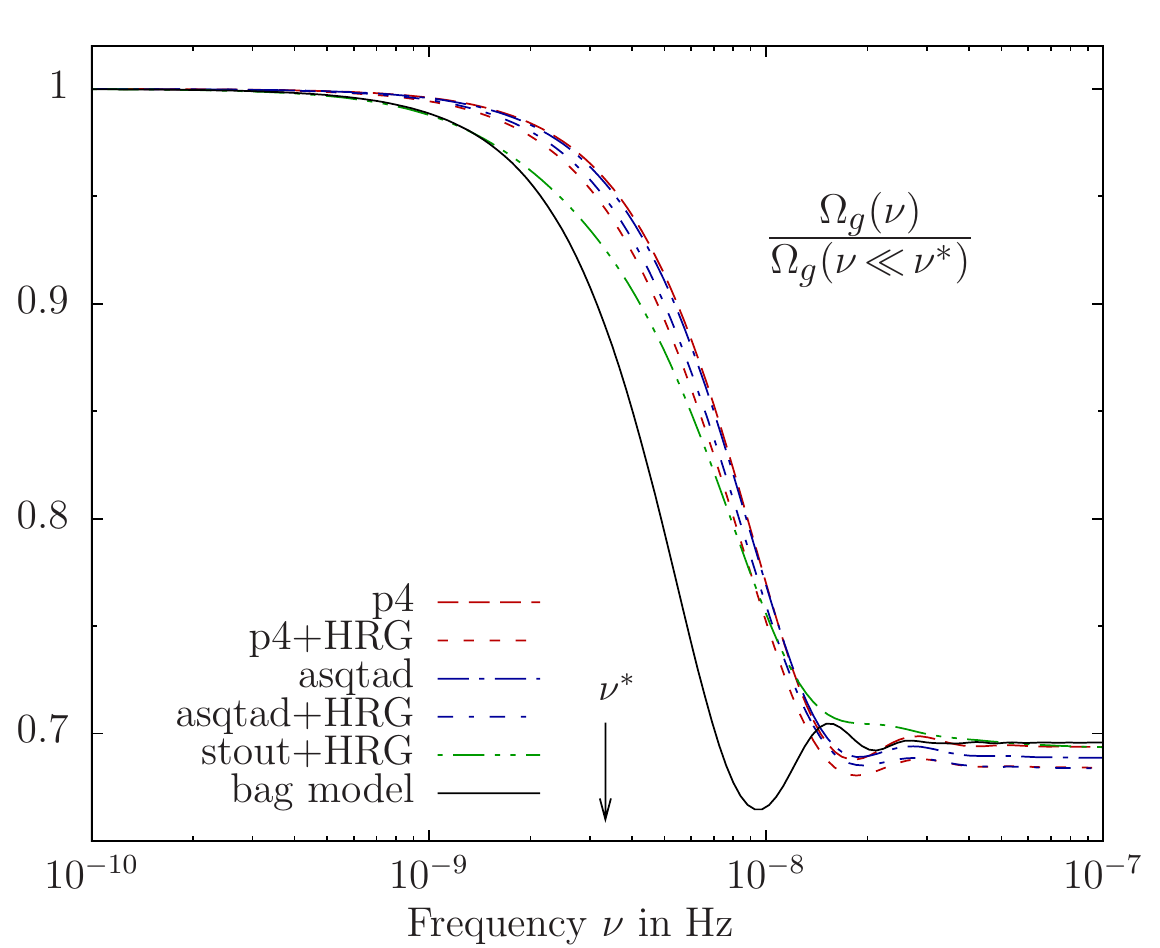}}
	\vspace{-0.5cm}

	\caption{(Color online) The energy density of GWs per logarithmic frequency interval after different types of QCD phase transition. Modes with high frequency are damped by a factor of $\sim0.7$ compared to low frequency modes. The solid line represents the result of a bag model calculation with $T_\textnormal{c} = 180$~MeV.  The dashed curves are calculated using lattice data. In three cases also the hadron resonance gas ansatz has been taken into account.}
	\label{fig:SpectrumLattice}
\end{figure}
This is, because in the bag model calculation we have assumed $g_1 = 51.25$ and $g_2 = 17.25$ degrees of freedom before and after the transition, respectively. Doing so we have taken the strange quarks to be non-relativistic and the pions to be fully relativistic. However, strange quarks and pions have masses of about the temperature scale at the transition, so we certainly oversimplified the situation with our assumptions. 

This discussion reveals that only if we restrict ourselves to calculate solely the effect of the QCD phase transition with given $g_1$ and $g_2$, the step-size is given by 0.7. If we include the evolution before and after the transition, every loss of relativistic degrees of freedom heightens the step. For example, including relativistic strange quarks and letting the pions become completely non-relativistic leads to a step of size
\begin{equation*}
	\left(\frac{g_2^{ \textnormal{without} \; \pi}}{g_1^{ \textnormal{with} \; s}}\right)^{1/3} = \left(\frac{14.25}{61.75}\right)^{1/3} = 0.613.
\end{equation*}
These extreme cases are chosen in Figures \ref{fig:PressureLattice}  and \ref{fig:RhoLattice} where we see that the lattice results lie in between. Note that in these diagrams only strongly interacting particles are considered. The total energy density and pressure are bigger by the amount of photons and leptons.

After all we should state that the step-size calculated with lattice data depends on the temperature of the system at the time when the calculation ends ($T_2$). The displayed p4 and asqtad curves are obtained for $T_2\approx$ 50$\,$MeV. As $T_2$ is varied from 10 MeV to 100 MeV, the damping factor sweeps out a range of less than one percent (for the p4 and asqtad parameterizations) and of about two percent, respectively (for the p4+HRG and asqtad+HRG parameterizations). 
Unfortunately, the interpretation of our findings is not quite as straightforward as claimed above: The parameterizations for the trace anomaly in \cite{PhysRevD.80.014504} diverge for $T \rightarrow 0$ and certainly lose reliability below 50$\,$MeV. Maybe this is the case well above that temperature: Even in the HRG case, no calculation at temperatures $T < 100\,$MeV is taken into account. At 100$\,$MeV, though, the trace anomaly is still not zero, which implies that strong interactions are not negligible. For lack of data in this low temperature region, we decided to extrapolate the given parameterizations to T = 0 using a simple power law dependence. At least, this guarantees radiation domination far away from the QCD scale. 

The ``stout'' parameterization reflects the HRG results also for $T < 100$ MeV and yields reliable results down to zero temperature. In this case, we ended our calculation at $T_2 = 10$ MeV. (The step-size stays the same at least down to $T_2 = 1$ MeV; however, the computing effort for a whole spectrum would have been much bigger.)

\subsection{Inflationary Scenario}\label{ss:Inflation}
\begin{figure}[t]
	\resizebox{8.8cm}{!}{	\includegraphics{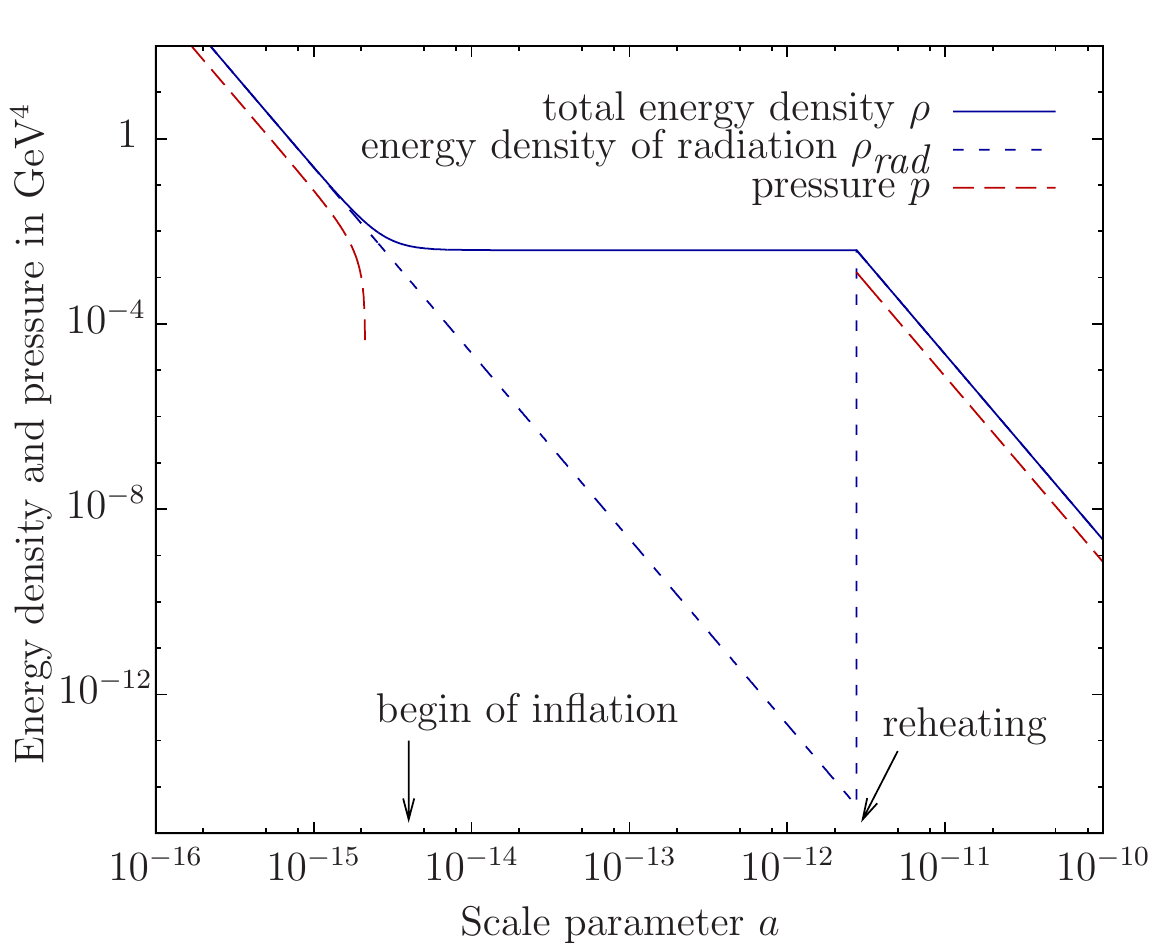}}
	\vspace{-0.5cm}
	\caption{(Color online) Energy density and pressure during a QCD phase transition with a short period of inflation. The total energy density decreases proportional to $a^{-4}$ until the bag constant $B \approx (250\, \textnormal{MeV})^4$ is no longer negligible. When the vacuum contribution is dominant, the total energy density does not depend on the scalefactor $a$. The same is true for the pressure which is negative during inflation: $p = -B$. The radiative energy density becomes negligible soon after inflation begins.}
	\label{fig:PAndRhoAtInfl}
\end{figure}

Since, until now, we have no field theoretical implementation of the inflationary scenario at hand, we return to the bag model and introduce the important features by hand: We start with the same temperature dependence of energy density and pressure as in the bag model calculation,
\begin{equation*}
	\rho(T) = \frac{\pi^2}{30}g_1 T^4 + B, \quad \quad p(T) =  \frac{\pi^2}{90}g_1 T^4 - B,
\end{equation*}
and let the system evolve without undergoing a transition. In this way we account for the demand that it is trapped in a false vacuum which is characteristic for a first order phase transition. Below the temperature
\begin{equation*}
	T_ \textnormal{infl} = \left( \frac{30}{g_1\pi^2} B \right)^{1/4}
\end{equation*}
a period of inflation sets in. During this time the temperature falls proportional to $1/a$, further supercooling the system. After some e-folds, the thermal part of the energy density is negligibly small in comparison with the vacuum contribution $B$ which we then force to decay into the conventional particle spectrum at about 160$\,$MeV: As in the bag model calculation, we take into account 
$g_2 = 17.25$ relativistic degrees of freedom. The temperature after this reheating is given by the bag constant and the degrees of freedom,
\begin{equation*}
	T_ \textnormal{r} = \left( \frac{30}{g_2\pi^2} B \right)^{1/4}.
\end{equation*}
It lies above the temperature when inflation started, but below the critical temperature $T_ \textnormal{c} $:
\begin{equation*}
		T_\textnormal{r} = \left( \frac{30}{g_2\pi^2} B \right)^{1/4} < \left( \frac{90}{(g_1 - g_2)\pi^2} B \right)^{1/4} = T_ \textnormal{c} ,
\end{equation*}
which is important if we want to avoid a transition back to the deconfined phase. The last statement holds true if $g_2 > g_1/4$. This relation is fulfilled even if the pions are not included after the transition.  In our case, the numerical value of $T_ \textnormal{r} $ is 162$\,$MeV. After reheating we let the system evolve without further manual input. Figure \ref{fig:PAndRhoAtInfl} illustrates the evolution of energy density and pressure during the subsequent periods described above.

\begin{figure}[t]
\resizebox{8.8cm}{!}{	\includegraphics{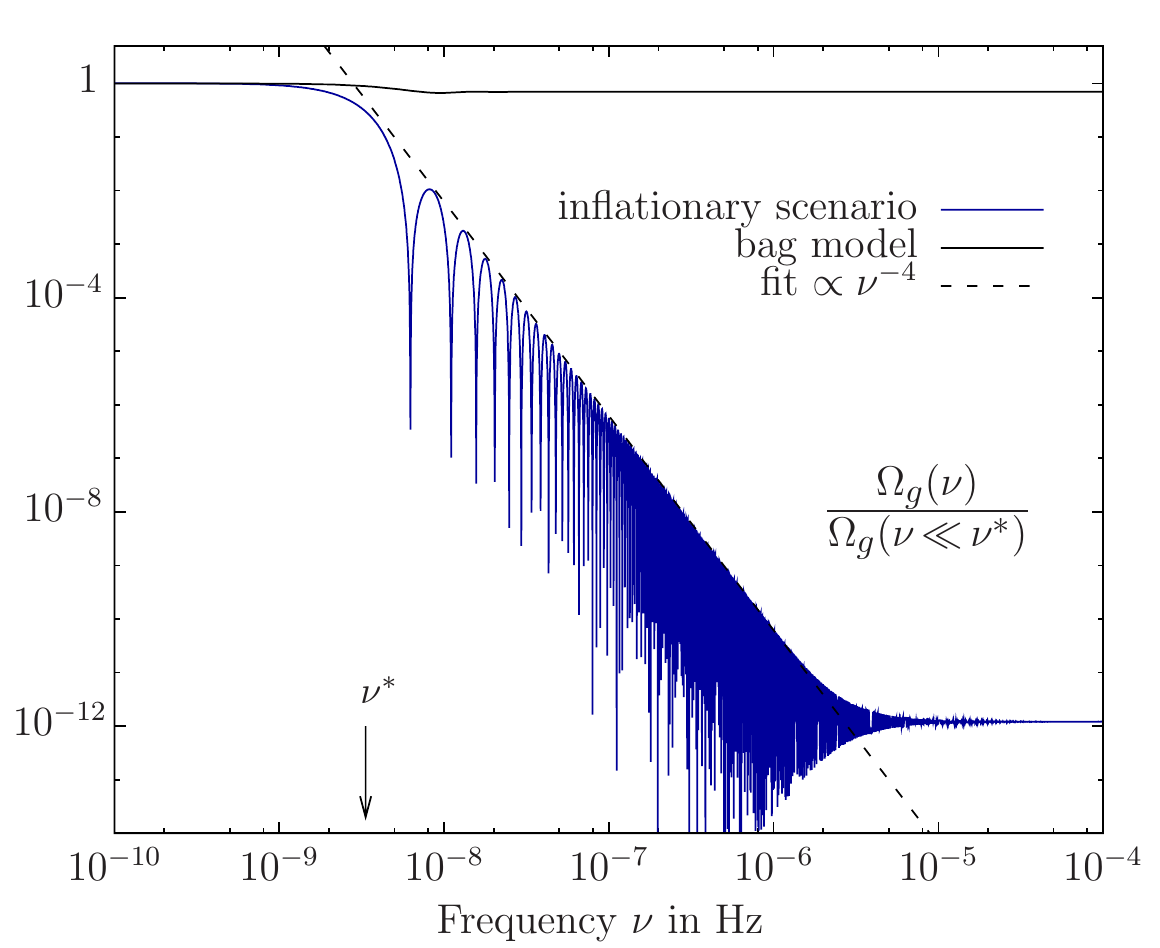}}
	\vspace{-0.5cm}
	\caption{(Color online) The energy spectrum of GWs after a QCD phase transition with a short period of inflation. The inflationary phase leads to a much stronger dilution than an ordinary bag model transition.}
	\label{fig:SpectrumInfl}
\end{figure}
The result of this procedure can be seen in Figure \ref{fig:SpectrumInfl}. The inflationary period dilutes the energy density in GWs much more than an ordinary phase transition. 

For modes with $\nu < 10^{-9}\,\Hz$ or $\nu > 10^{-5}\:\Hz$, $\Omega_ \textnormal{g} (\nu)$ is constant. This is because the modes with $\nu < 10^{-9}\,\Hz$ stay outside the horizon until the radiation dominated era after the little inflation, whereas the modes with $\nu > 10^{-5}\,\Hz$ enter the horizon already before the little inflation and stay inside until the end of the calculation.

The modes with $10^{-9}\,\Hz < \nu < 10^{-8}\,\Hz$ have wavelengths of order the Hubble horizon when the short period of inflation begins. They slightly dip into the horizon and are the first to be driven out again where they experience no damping anymore. So they are much less suppressed than the modes with higher frequencies that are already well within the horizon when inflation begins and need more time to reach superhorizon scales. During our calculation, the intermediate modes with frequencies of roughly $5\:\cdot10^{-9}\,\Hz < \nu < 5\:\cdot10^{-6}\,\Hz$ cross the horizon three times: We set the initial conditions when all modes have wavelengths $\lambda > 1/H$; then the intermediate and the high frequency modes enter the horizon in the radiative universe before the QCD transition; the short period of inflation stretches the intermediate modes again to superhorizon scales. After reheating, the universe is radiation dominated again, letting the Hubble horizon grow faster than the wavelengths: The calculation is continued until all modes are inside the horizon again. In order to estimate the final shape of the power spectrum $\mathcal P_ \textnormal{g}$ at intermediate frequencies, the effects of all three horizon crossings must be taken into account:
\begin{equation}
\label{sequenceRadInflRad}
	\mathcal P_ \textnormal{super} ^{(1)} \xrightarrow{\textnormal{radiation}}
	\mathcal P_ \textnormal{sub} ^{(1)} \xrightarrow{\textnormal{inflation}}
	\mathcal P_ \textnormal{super} ^{(2)} \xrightarrow{\textnormal{radiation}}
	\mathcal P_ \textnormal{sub} ^{(2)}.
\end{equation}

After the thermal contributions to $\rho$ and $p$ have become small compared to the bag constant $B$, the equation of state of the medium is $\;p = -\rho$. This leads to
\begin{equation*}
	a_ \textnormal{ex}(k) = \frac{k}{H_ \textnormal{ex}(k)} \propto k	
\end{equation*}
during the little inflation. We denote the scale parameter at the first horizon entry with $a^{(1)}_ \textnormal{in} $ and the one at the second horizon entry with $a^{(2)}_ \textnormal{in} $ and calculate
\begin{align}
	\mathcal P_ \textnormal{sub} ^{(2)}(k,a) &= \mathcal P_ \textnormal{super} ^{(2)}(k) \left( \frac{a_ \textnormal{in} ^{(2)}(k)}{a} \right)^2 \notag\\
	&= \mathcal P_ \textnormal{sub} ^{(1)}(k,a_ \textnormal{infl} ) \left( \frac{a_ \textnormal{infl} }{a_ \textnormal{ex} (k)} \right)^2 \left( \frac{a_ \textnormal{in} ^{(2)}(k)}{a} \right)^2\notag\\
	&= \mathcal P_ \textnormal{super} ^{(1)}(k) \left( \frac{a_ \textnormal{in} ^{(1)}(k)}{a_ \textnormal{infl} } \frac{a_ \textnormal{infl} }{a_ \textnormal{ex} (k)} \frac{a_ \textnormal{in} ^{(2)}(k)}{a} \right)^2\label{PsubPsuper}\\	
	&= \mathcal P_ \textnormal{super} ^{(1)}(k) \frac{1}{a^2} \frac{1}{k^6} \propto \frac{1}{k^6}.\notag
\end{align}
With $a_ \textnormal{infl} $ we denote any scale parameter for which the considered mode is still inside the horizon after the first horizon entry.\\
Since $\Omega_ \textnormal{g} (k) := k^2 \mathcal P_ \textnormal{g} (k)$, we obtain
\begin{equation*}
	\frac{d \ln \Omega_ \textnormal{g} (\nu)}{d \ln \nu} = -4,
\end{equation*}
which is in agreement with our numerical result plotted in Figure \ref{fig:SpectrumInfl}. In our calculation inflation lasts until the scale factor has grown by a factor of 1000. Since the Hubble parameter $H$ is constant during that time and the physical wavelength of the modes is stretched proportional to $a$, the modes being driven out the horizon have a frequency range of three orders of magnitude. This leads to an overall damping of the high frequency modes in $\Omega_ \textnormal{g} $ by a factor of $(10^{3})^{-4} = 10^{-12}$, affirming again the numerical outcome. This result can also be obtained by noting that
\begin{equation}
\label{OmegaAndA}
	\Omega_ \textnormal{g}  (a) \propto a^{-4}
\end{equation}
because gravitational radiation is redshifted in the same way as light.

\begin{figure}[t]
\resizebox{8.8cm}{!}{	\includegraphics{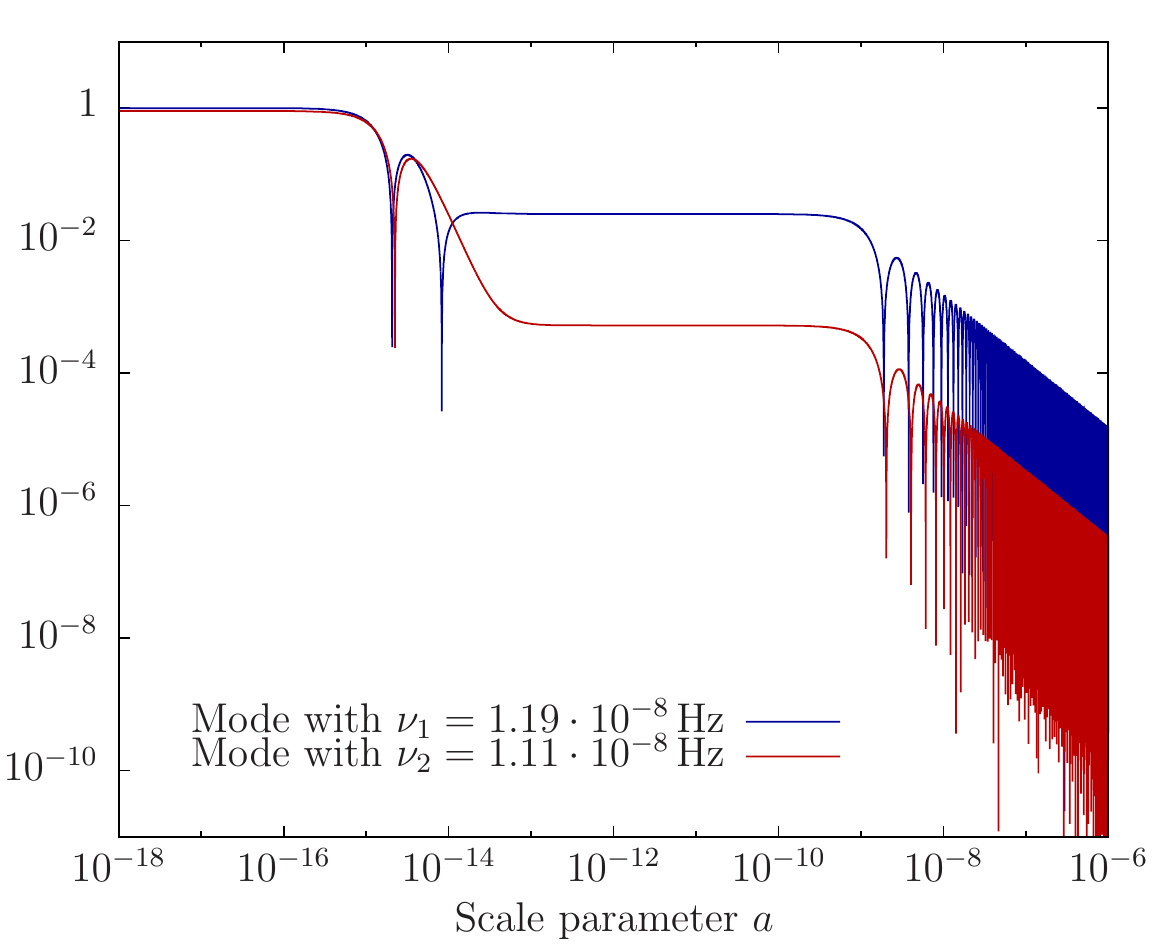}}
	\vspace{-0.5cm}
	\caption{(Color online) Temporal evolution of two modes during a QCD phase transition with a period of inflation. The mode with frequency $\nu_1$/$\nu_2$ is at maximal/small displacement when leaving the horizon. Therefore, the former appears as a maximum in the spectrum and the latter as a minimum.
	}
	\label{fig:TwoModesDuringInfl}
\end{figure}

Now we should discuss the origin of the oscillations in $\Omega_ \textnormal{g} $ for the modes with intermediate frequencies: When a mode exits the horizon, it stops oscillating. If this occurs while $h_{\bf k}(t) \approx 0$, the mode is frozen at a very small value. If this occurs while $h_{\bf k}(t)$ has a maximum, the mode is fixed at a large value. This phase effect is similar to the one discussed in \cite{Schmid1996qd} for scalar perturbations.
 We illustrate the effect of an inflation on GWs in Figure \ref{fig:TwoModesDuringInfl}. We expect a related mechanism to cause the slight oscillation in the spectrum after a normal bag model phase transition (see Figure \ref{fig:SpectrumLattice}). 

\subsection{Towards a Field Theoretical Implementation}\label{ss:FieldTheory}
We would like to extend our studies of an inflationary scenario for the QCD phase transition: In order to approach our aim of a field theoretical implementation, we introduce a scalar field $\chi$ whose potential energy replaces the bag constant. During the transition, the field passes from a false vacuum state into the true vacuum with zero potential energy. This corresponds to setting the vacuum energy density $B$ after the transition to zero.

\begin{figure}[t]
\resizebox{8.8cm}{!}{	\includegraphics{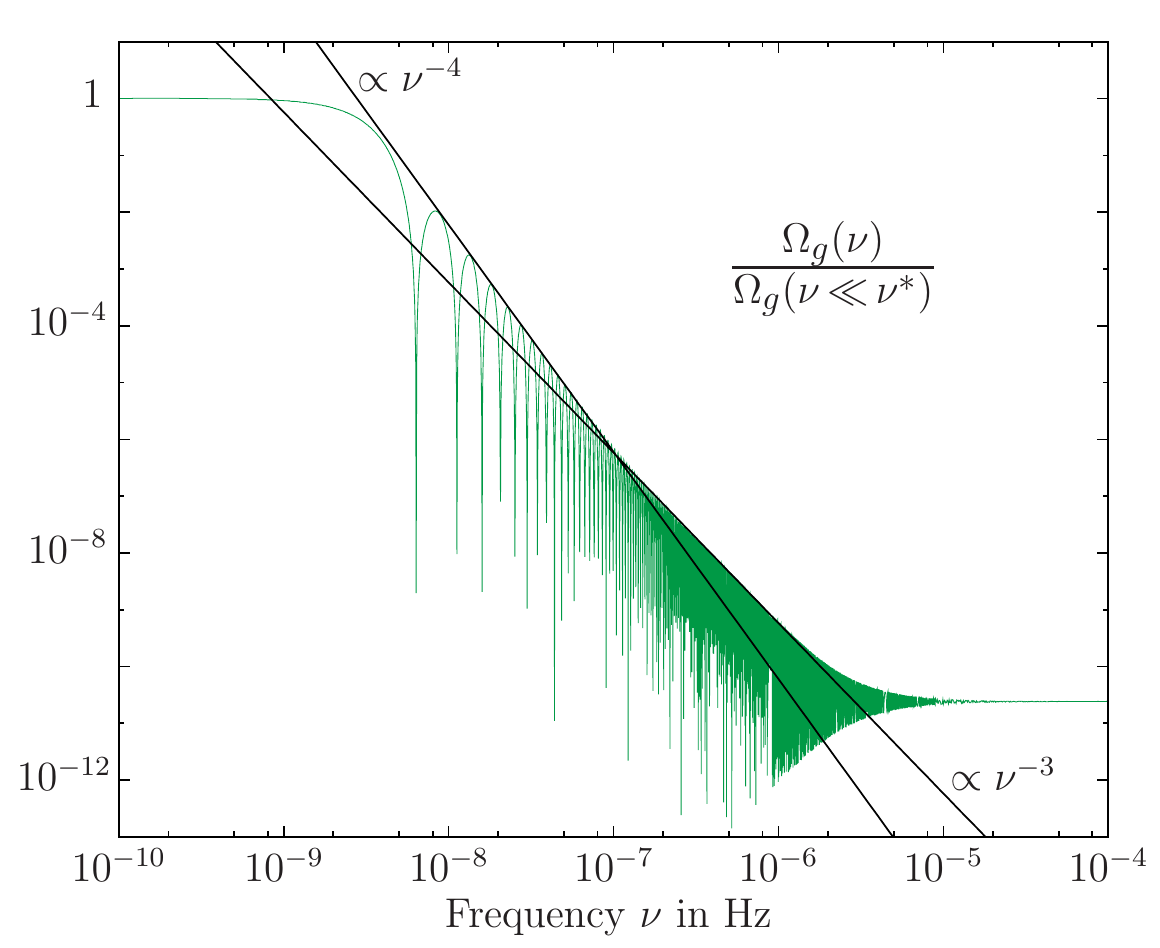}}
	\vspace{-0.5cm}
	\caption{(Color online) The energy spectrum of GWs after a QCD phase transition including a little inflation followed by a period which is dominated by the kinetic energy density of a scalar field. The spectrum of modes which reenter the horizon during this stage falls like $\nu^{-3}$, whereas the energy density in modes which reenter during the subsequent radiation dominated epoch is proportional to $\nu^{-4}$.}
	\label{fig:InflChi}
\end{figure}

\begin{figure}[t]
\resizebox{8.8cm}{!}{	\includegraphics{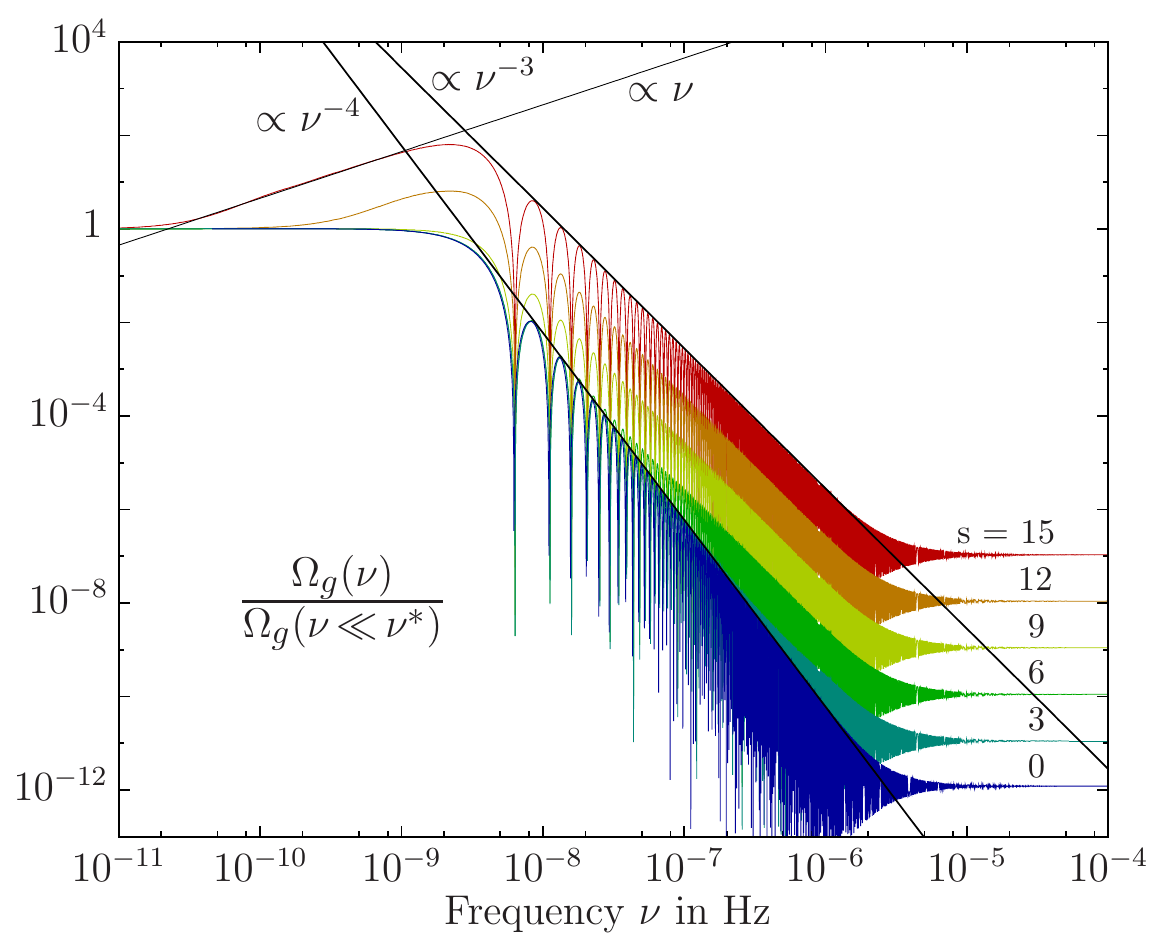}}
	\vspace{-0.5cm}	\caption{(Color online) Energy spectra of relic GWs after different types of QCD transition. The parameter $s$ indicates the duration of the oscillation dominated period within the corresponding scenario. In the case $s=0$ there is no oscillation domination included.}
	\label{fig:InflChiManyModes}
\end{figure}

We calculate within a strongly simplified model: 
Before the transition, we take the dilaton field to be constant at a value associated with high potential energy, i.e. we calculate within the bag model. After the short period of inflation, we let the potential energy transform into kinetic energy (kination). Doing so, we assume a transition to a state where the dilaton oscillates within a potential well with negligible potential energy.
Since
\begin{equation*}
	\rho = \frac{1}{2} \dot\chi^2 + V(\chi) \quad\quad \text{and} \quad\quad p = \frac{1}{2} \dot\chi^2 - V(\chi),
\end{equation*}
we get an equation of state with $w=1 \; \Rightarrow \; \rho \propto a^{-6}$. However, most probably the oscillating dilaton field is suppressed even much faster because, during preheating, it decays into other fields on timescales which are very small compared to the Hubble time during the QCD era ($\sim10^{-5}$ seconds). In this case, the ``oscillation dominated'' period would last too short a time in order to be discernible in the GW spectrum. In our calculation, we assume that the oscillation of the dilaton is only damped by Hubble expansion over a period of time which is long enough to be resolved in the spectrum. Afterwards, the remaining energy is converted into a suitable spectrum of relativistic particles, in the same way as it has been done within the previous calculation. This scenario allows us to get a better feeling for the signature of a sequence of periods dominated by different kinds of energy densities. 
Since, once more, reheating is put in by hand, we can freely decide when it should occur. The measure will be the ratio by which the energy density of oscillation is diluted.
The sequence \eqref{sequenceRadInflRad} is now modified as follows:
\begin{equation*}
	\mathcal P_ \textnormal{super} ^{(1)} \xrightarrow{\textnormal{radiation}}
	\mathcal P_ \textnormal{sub} ^{(1)} \xrightarrow{\textnormal{inflation}}
	\mathcal P_ \textnormal{super} ^{(2)} \xrightarrow[\textnormal{or radiation}]{\textnormal{oscillation}}
	\mathcal P_\textnormal{sub} ^{(2)}.
\end{equation*}	
From equation \eqref{ainAndk} we have
\begin{equation}
\label{aoscOfk}
	a^ \textnormal{osc} _ \textnormal{in}  \propto \frac{1}{\sqrt{k}}
\end{equation}
during oscillation domination. This leads to a power spectrum
\begin{align*}
	\mathcal P_ \textnormal{sub} ^{(2)}(k,a)
	=  \mathcal P_ \textnormal{super} ^{(1)}(k) \frac{1}{a^2} \frac{1}{k^5}
\end{align*}
as can be seen from equation \eqref{PsubPsuper}. We therefore expect the energy spectrum to be proportional to $\nu^{-3}$. If the oscillation dominated period lasts until the modes with frequencies $\nu \lesssim 10^{-9}\, \Hz$ enter the horizon, the shape of the low frequency spectrum follows from equation \eqref{OmegaAndW}: $\Omega_ \textnormal{g} (\nu) \propto \nu$. These expectations are confirmed by our numerical results shown in Figures \ref{fig:InflChi} and \ref{fig:InflChiManyModes}.
Figure \ref{fig:InflChi} displays the spectrum of primordial GWs after an inflationary QCD phase transition with an oscillation dominated period. It lasts until the energy density is diluted by a factor $10^{-4} \equiv 10^{-s}$, where, for later use, we have defined the parameter $s$ to be the negative decadic logarithm of this factor.
Correspondingly, the scale parameter grows by a factor $10^{s/6} = 10^{2/3}$. As can be seen from equation \eqref{aoscOfk}, the modes which enter the horizon during this time cover an interval corresponding to a factor $10^{s/3} = 10^{4/3}$. In Figure \ref{fig:InflChi}, this interval roughly lies between  $10^{-7}\:\Hz$ and $3\cdot10^{-4}\:\Hz$, where the spectrum shows the expected behavior. The second horizon entry of the lower frequencies happens during radiation domination and, consequently, the resulting spectrum is proportional to $\nu^{-4}$.
In Figure \ref{fig:InflChiManyModes} we compare GW spectra corresponding to scenarios with various durations of the oscillation dominated period. The result without oscillatory period is also displayed. The five further curves result from computations including oscillation dominated periods with durations given by the respective value of $s$. In the cases $s = 12$ and $s = 15$ we also see the behavior $\Omega_ \textnormal{g} (\nu)\propto\nu$ for the modes which did not enter the horizon before the little inflation. 
This is consistent with the analytical estimate \eqref{OmegaAndW} which was obtained from energy-momentum conservation and the fact that GWs decay as $1/a$ after horizon entry. The slope of the spectrum is positive, because the rate of entering modes is higher than during radiation domination: This leads to a stronger damping of low frequency modes since they enter the horizon earlier.

\section{Conclusions and Outlook}\label{s:Conclusions}

We have considered the gravitational background radiation assuming a flat spectrum before the cosmological QCD phase transition and calculated the spectrum afterwards. A few different scenarios have been taken into consideration: First we calculated within the bag model; then we made use of lattice data for zero chemical potential; finally, we explored the consequences of a QCD transition including a short period of inflation. 

We have seen that the loss in relativistic degrees of freedom imprints a step in the energy density spectrum of GWs. In the case of negligible entropy production during the transition, the step-size is determined by the ratio of degrees of freedom in the two phases. In contrast, the large entropy release within the inflationary model accompanies a very different shape of the spectrum: Inflation much more attenuates the energy density of modes inside the horizon than it is expected in case of entropy conservation. 

We now work at field theoretical models which imply the inflationary scenario \cite{BoeckelToAppear}. Within such a model we could calculate the evolution of energy density and trace anomaly in order to use it in the computation of the resulting spectrum. But there is an even more important purpose: A consistent model within the framework of effective field theory certainly increases the reliability of the scenario. For example, an important point addressed in \cite{BoeckelToAppear} is whether it is possible to trap the system until such time as inflation has lasted long enough to dilute the baryon density down to the numbers required by observation.  

\begin{figure}
\resizebox{8.8cm}{!}{	\includegraphics{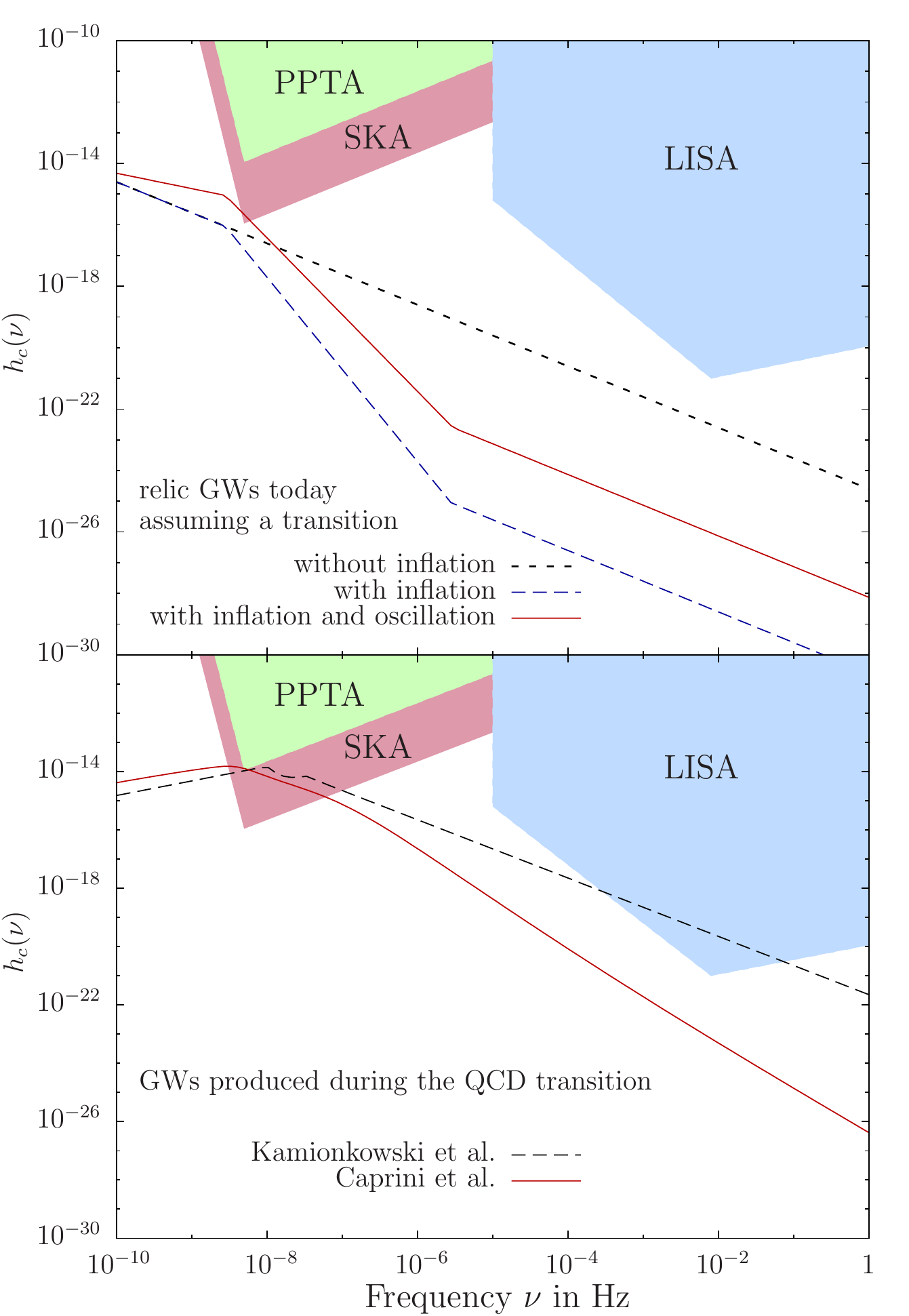}}
	\vspace{-0.46cm}
	\caption{(Color online) GWs after the QCD transition and their detectability. The (predicted) sensitivities of three GW detectors are displayed. PPTA (Parkes Pulsar Timing Array) already constrains the GW production during the QCD phase transition. The corresponding spectrum is displayed on the basis of \cite{PhysRevD.49.2837} (Kamionkowski et al.) and \cite{PhysRevD.82.063511} (Caprini et al.). See also \cite{2004NewAR..48..993K, 2008PhRvD..78d3003K}. For the amplitude of relic GWs, we use the largest amplitude consistent with the COBE constraint given in \cite{1997stgr.proc....3A}. The solid line in the upper panel sketches the result of a calculation including a little inflation and a subsequent oscillation dominated period with $s=15$.}
	\vspace{-0.2cm}
\label{figDetectors}	
\end{figure}

We neglect the possible production of perturbations by amplification of vacuum fluctuations. This is because their spectrum is suppressed as $(\Lambda_\textnormal{QCD}/m_\textnormal{Pl})^2$ with respect to a corresponding spectrum created at energies around the Planck scale.
But an important point is the GW production during the phase transition by bubble collisions and subsequent turbulences within the plasma.

In Figure \ref{figDetectors}, we have displayed possible GW spectra after a first-order QCD phase transition. 
The characteristic amplitude $h_ \textnormal{c} $ is defined from the energy density per logarithmic frequency interval (which we previously used) as
\begin{equation*}
	h_ \textnormal{c} (\nu) := \frac{0.9 \cdot 10^{-18}\, \Hz}{\nu} \frac{h}{0.7} \sqrt{\Omega_ \textnormal{g} (\nu)},
\end{equation*}
where $h$ is the Hubble parameter today in units of $100\, \text{km}/(\text{s} \cdot \text{Mpc})$.
The upper panel shows the characteristic amplitude of GWs generated during inflation and their spectrum after an additional inflationary period at QCD time. Likewise, the result of a calculation including both, inflation and oscillation domination is presented. Assuming an extreme value of $s=15$, we might hope that even the spectrum of primordial GWs could be measured by SKA. However, a scenario with a dilaton field being stable for more than a Hubble time seems to be far from realistic.

The GWs which are produced during the transition can be seen in the lower panel of Figure \ref{figDetectors}. The dominant contributions come from bubble collisions and from the turbulences thereafter. The spectra displayed here are obtained from parameterizations given in \cite{PhysRevD.82.063511,PhysRevD.49.2837}. They show the maximal GW production consistent with PPTA measurements. Reference \cite{PhysRevD.82.063511} assumes the presence of magnetic fields in the turbulent plasma. The most important deviation between the spectra is the different behavior in the high frequency regime:  \cite{PhysRevD.49.2837} includes multi-bubble collisions and therefore obtains a flatter spectrum $\propto \nu^{-1}$ for high frequencies. This result is also confirmed in the more recent work \cite{2008JCAP...09..022H} where, however, turbulences are not accounted for. Neither one of the parameterizations includes an inflationary period. A corresponding spectrum still has to be calculated. 

We also display the sensitivity of three relevant GW detectors \cite{2004NewAR..48..993K}: PPTA (Parkes Pulsar Timing Array) is already running and has constrained the spectrum emanating from the transition. SKA (Square Kilometre Array) and LISA (Laser Interferometer Space Antenna) will further push down the limits for GW production at QCD time. In the more remote future, even some features of the shape of the relic spectrum could be revealed by combining measurements of Planck \cite{Planck2005} at $\nu \sim 10^{-18}\, \Hz$ and of the proposed Big Bang Observer (BBO \cite{2005gr.qc....12039C}) at $\nu \sim 1 \,\Hz$. Given that the Planck mission succeeds in finding imprints of relic GWs, BBO could potentially discriminate between the standard QCD cross-over and an inflationary phase transition.
\vspace{-0.02cm}

\section*{Acknowledgments}
\vspace{-0.03cm}
We would like to thank Ruth Durrer and Dominik Schwarz for useful discussions.
This work is supported by BMBF under grant FKZ 06HD9127, by the German Research Foundation
(DFG) within the framework of the excellence initiative through the Heidelberg
Graduate School of Fundamental Physics, the Gesellschaft f\"ur Schwerionenforschung
GSI Darmstadt, the Helmholtz Graduate School for Heavy-Ion Research (HGS-HIRe), the Graduate
Program for Hadron and Ion Research (GP-HIR) and the Alliance Program of the Helmholtz Association
(HA216/EMMI). Simon Schettler acknowledges support by the IMPRS for Precision Tests of Fundamental Symmetries.

\end{document}